# Raman Study of CFC Tiles Extracted From the Toroidal Pump Limiter of Tore Supra

C. Pardanaud[a,*], G. Giacometti[a], C. Martin[a], R. Ruffe[a], T. Angot[a], E. Aréou[a], B. Pégourié[b], E. Tsitrone[b], T. Dittmar[b], C. Hopf[c], W. Jacob[c], T. Schwarz-Selinger[c], P. Roubin[a]

[a]*PIIM, CNRS - Université de Provence, Centre Saint Jérôme, 13397 Marseille cedex 20, France.*
[b]*CEA, IRFM, 13108 Saint-Paul-lez-Durance, France.*
[c]*MPI für Plasmaphysik, EURATOM Association, Boltzmannstr. 2, 85748 Garching, Germany.*

**Abstract**

The structure of six tiles extracted from the erosion and deposition zones (thin and thick deposition) of the Tore Supra toroidal pump limiter (TPL) have been analysed in the framework of the DITS campaign using micro-Raman spectroscopy. This post-mortem analysis gives information on both carbon structure and D content. We have found that the carbon structure is most often similar to that of plasma-deposited hard amorphous carbon layers. The role of the surface temperature during the discharge in the D content is investigated: in all locations where the temperature does not reach more than 500°C the D content seems to be roughly uniform with D/D+C ≈ 20%.

# 1. Introduction

Understanding and controlling fuel retention is a crucial issue for fusion. The Deuterium Inventory in Tore Supra (DITS) project aims at a better understanding of retention mechanisms due to either formation of carbon layers at the surface of plasma-facing components (PFCs) or migration into the bulk of PFCs. A plasma campaign was dedicated to the D inventory through a detailed particle balance and after that a sector of the Toroidal Pump Limiter (TPL) was dismantled for extensive post-mortem analyses [1]. The TPL surface, a castellated structure of carbon fiber composite (CFC) tiles (five surfaces for each tile: the top of the tile and the four sides into the gaps between the tiles), showed three distinct zones dominated either by erosion or by thin or thick deposition [2].

Particle balance analysis showed that 10.4 g of D were loaded in the machine at the end of the campaign. 4.85 g were found by post mortem analysis (NRA, TDS) on the TPL: 43 % of that amount in the erosion zone, 37 % in the thin deposit zone and the remaining 20 % in the thick deposit zone [1]. The important contribution to D retention of deposition inside the gaps was also evidenced, the major contribution (33 % of the total amount) coming from the erosion zone.

From thermography measurements [1] the following temperatures were deduced: ~ 120 °C for both top and gap surfaces of tiles situated in the thin deposit region, ~ 200 °C for top surfaces of eroded tiles, and ~ 500 °C for gap surfaces of both eroded and thick deposit tiles. For the top surface of thick deposit tiles T varied between 500 and 900 °C, non uniformly across the tile.

We have analyzed the surfaces of six tiles extracted from the three different zones of the TPL (eroded, thick and thin deposit zones, labeled according to ref. [2]) using micro-Raman spectroscopy. Results are compared to those obtained for a heat treated reference sample (a plasma-deposited hard amorphous carbon layer [3]) to get information about the correlation between temperature treatment and Raman signature. Note that Raman analysis could be a fast way to estimate the D content of carbon layers (depth probed: from a few tens to a few hundreds of nanometers) in tokamaks.

This paper is organized as follows: part 2 gives specificities of the micro-Raman spectroscopy, in parts 3 and 4, carbon structure and D content are deduced by comparison with the reference sample, discussion and conclusion are given in part 5.

## 2. Micro-Raman spectrometry

Micro-Raman spectroscopy, already used for characterizing carbon deposits in tokamaks [4-7], is a non destructive technique routinely used to characterize different forms of C-based materials: diamond, graphite, nano crystalline graphite, amorphous carbons (a-C, a-C:H), etc. Interpreting the 1000-1800 cm$^{-1}$ spectral region gives information on structural and chemical (hybridization of the carbon atoms and H, D content) properties [8-10]. This technique probes the sample surface with a ~ μm$^2$ spot and the information depth ranges from a few tens to a few hundreds of nanometers [11]. Because of the spot size and because of the fastness of this technique (1 spectrum typically requires ~100 s, with a good signal to noise ratio), it was used here to test the homogeneity of samples at a centimetre scale by randomly recording a large number of spectra. Raman spectra were obtained using a Horiba-Jobin-Yvon HR LabRAM apparatus (laser wavelength: $\lambda_L$=514.5 nm, 100X objective, resolution ~1 cm$^{-1}$). The laser power was chosen at P ~ 2 or ~ 0.2 mW m$^{-2}$ to have a good signal/noise ratio and/or to prevent damages. To check repeatability, spectra were recorded several times at the same spot, or in close vicinity.

Figure 1 displays two Raman spectra in the 1000-1800 cm$^{-1}$ spectral region which were recorded on the top surface of a tile situated in the thick deposit region and on the reference plasma-deposited hard a-C:H layer. The a-C:H layer is a typical hard hydrocarbon film with a hydrogen content of 0.33 and density of about 1.7 g cm$^{-3}$ (see, e. g., Table 1 in [3], data for methane at -200V bias). Both Raman spectra are similar, composed of two broad bands: the D and G bands, plus a linear background. It is known that these bands are caused by sp$^2$ carbon atoms: the G band is due to the bond stretching of both aromatic and aliphatic C-C pairs, whereas the D band is due to the breathing of aromatic rings and/or the presence of disorder [10, 12]. The sp$^3$ carbons do not give rise to a spectral contribution on the spectra with $\lambda_L$=514.5 nm but can influence band positions and widths depending on the coupling with $sp^2$ carbon atoms. It is also known that the G band wavenumber, $\sigma_G$, together with its broadening, $\Gamma_G$, allows estimating the size of the aromatic domains contained in the material (local order increasing with this size) [8]. For a-C:H layers, the linear background, with a slope $\eta$, is due to photoluminescence. The ratio of $\eta$ to the G band intensity, $I_G$, is related to the D content and our determination of the D content using Raman micro-spectroscopy is based on the empirical relation of ref. [8], where H/C = 21.7 + 16.6 Log [$\eta/I_G$ (μm)].. The Raman parameters $\sigma_G$, $\Gamma_G$ and $\eta/I_G$ were measured directly on the spectra as indicated in Fig. 1, preventing from ambiguousness due to a model-dependent fitting procedure.

### 3. Carbon structure

Figure 2 plots $\sigma_G$ against $\Gamma_G$ for the top and the gap surfaces of the three kind of tiles and for the reference sample, a heat treated a-C:H layer. Each point in this figure corresponds to a spectrum. For the heat treated a-C:H layer, the relation between the two Raman parameters is nearly linear, points slightly deviating from a straight line only for $\Gamma_G$ from 120 to 60 cm$^{-1}$. $\sigma_G$ increases from 1520 cm$^{-1}$ for the as-deposited sample to 1600 cm$^{-1}$ for the sample heat-treated at ~ 450 °C, whereas the corresponding $\Gamma_G$ decreases from 200 cm$^{-1}$ to 60 cm$^{-1}$. Under heating, the size of aromatic domains contained in a-C:H layers, and then the local order, increases [9].

For the DITS samples, the data are spread around a similar relation, indicating that these samples are amorphous carbons. Samples with Raman parameters situated in the upper left corner of this figure are better ordered than samples situated in the lower right corner. For the top surface of the thin and thick deposit tiles, the structure is inhomogeneous across the tile: $\Gamma_G$ varies from ~ 90 to 190 cm$^{-1}$ for the former, and from ~ 60 to 190 cm$^{-1}$ for the latter. Note that the highest order ($\Gamma_G$ ~ 60 cm$^{-1}$) is observed for the top surface of the thick deposit tiles. On the contrary, for the top surface of the eroded tile, carbon is poorly organized, and relatively homogeneous across the tile with $\Gamma_G$ ~ 190 cm$^{-1}$. For the gaps of the thin and thick deposit tiles, inhomogenity is also evidenced and $\Gamma_G$ varies from ~ 90 to 200 cm$^{-1}$ for the former, and from ~ 80 to 200 cm$^{-1}$ for the latter. For the gaps of the eroded zone, $\Gamma_G$ varies from ~ 95 to 190 cm$^{-1}$ and contrary to the top surface, the carbon structure is inhomogeneous. Inside the gaps this inhomogenity is dependent on the depth: for the thick deposit tile there is a region with a local order near the top surface down to roughly ~1 mm (empty symbols in figure 2, depths in the gaps are estimated by measuring the distance from the top surface of the tile with the optical microscope used to do the Raman spectra), whereas no local order is measured down to ~3 mm in the gap (full symbols). For the tile from the erosion zone, the transition depth between locally ordered and less ordered amorphous carbons is situated at ~ 0.3 mm. For the thin deposit tile, we were not able to determine a transition depth because deposits are very few and flaky, but again, local order can be found near the top whereas only poorly ordered carbon deposits are found deeper in the gap.

An estimation of the density can also be made using results from the literature, ρ and $\Gamma_G$ being linearly related for as-deposited samples [8]. ρ can thus be estimated at ~ 2.0 ( ± 0.2) g cm$^{-3}$ for the less ordered samples. Unfortunately neither the role of the surface temperature nor the role of the heating temperature have been studied for a large variety of amorphous carbons, and density cannot be estimated here from spectra indicating a local order.

## 4. Estimation of the D content

Table 1 displays the D contents estimated in the thin deposit, thick deposit and eroded tiles using the empirical relation of ref. [8]. Except for the top surface of the thick deposit tile, the D contents obtained are similar for the top surfaces of the three zones, ranging between 13 and 28 %. A more refined analysis shows that the D content is inhomogeneous across tiles: it varies from 13 to 20 % for the F27T10 tile top surface for example. In addition, two tiles extracted from the same zone do not have exactly the same D content: on the top surface of thin deposits, D/D+C is slightly larger for the F17T7 tile (17-24%) than for the F10T10 tile (14-18%). Note that the top surface of the thick deposit zone is the only region where we can find $\eta/I_G$ ~ 0, i.e. no D.

Our results are in agreement with previous NRA and SIMS measurements from the top surfaces of thin and thick deposit tiles [5], for which these techniques give a D/D+C ratio of ~ 23 % and 20 %, respectively. Thick deposits can easily flake and this can explain some discrepancies and some inhomogeneities, the deposits probed possibly corresponding to campaigns older than the DITS campaign. In the case of the top surface of the eroded tile, a value of ~ 6% was measured by NRA and SIMS, much lower than that obtained with Raman (16 – 20 %). It has been shown by transmission electron microscopy that, for eroded tiles, CFC is homogeneously covered by an amorphous layer of only few tens of nm thick, which is due to the CFC amorphisation by ion bombardment and which is clearly distinct from deposits (the thickness of this layer is roughly three orders of magnitude lower than the thickness of deposits found in thin or thick deposition zones [13]). Deuterium thus probably mainly comes from this layer, and NRA or SIMS depth resolution (~ µm) can significantly lower the measured D content. Taking into account for the erosion zone an amorphous layer thickness of 30 nm, a 20 % D/D+C content, $\rho$ ~ 2.0 g cm$^{-3}$, and a 3.5 m$^2$ total surface leads to ~ 6 mg of D retained in the top surface of the tiles from erosion zone, i.e. ~ 0.1 % of the total D retention, and ~ 0.3 % of the D retention in the erosion zone. This confirm that the main contribution to retention coming from the top surface of eroded tiles is due to deposition inside porosities.

5. **Discussion**

We have studied six tiles exposed in Tore Supra and extracted from the TLP using Raman micro-spectroscopy and we have found that the carbon structure is mainly amorphous and deuterated. The D content is found similar everywhere (D/D+C ~ 20 %), except on the top surface of the thick deposit tile where D/D+C can be as low as 0 %. This is probably related to the high temperature that these latter deposits can reach (> 500 °C, 500 °C being the temperature at which hydrogen begins to desorb from amorphous layers, see [14]) due to their low thermal contact with the active cooling system of the TPL. This can also be correlated to our estimation of the carbon structure local ordering, the highest order (spectra with $\Gamma_G \approx 60$ cm$^{-1}$) being obtained in the case of the top surface of the thick deposit tile. Note that however, this surface is very inhomogeneous, the highest disorder being also found for these tiles (spectra with $\Gamma_G$ ~ 190 cm$^{-1}$). This lack of homogeneity can be explained by flaking of the deposits. This also shows that the discharge history (see operational conditions details of the DITS campaign in [2]) induces a complex thermal cycling of the deposits, significantly changing their structure.

The top of the thin deposit tiles is also heterogeneous, with the presence of carbon with local order. In this case, deposits are well-attached and the temperature remains low (T ~ 120°C). The local order can be due the intrinsic structure of the deposits resulting from deposition of aromatic species coming from the plasma phase [7]. We have also evidenced a gradient of order inside the gaps that can be due either to a thermal gradient or to different deposition mechanisms.

On the contrary, on the top of the erosion zone, the structure is remarkably homogeneous, with no local order: in this case the amorphous nature of carbon is most probably due to ion bombardment of CFC rather than to deposition.

## 6. Conclusion

As part of the DITS campaign, extensive Raman investigations have been performed on relevant regions of the toroïdal pump limiter of Tore Supra: the top and gap side surfaces of the erosion, thick and thin deposit zones.

The main results are (i) the D-content measurement (D/D+C~20%) and (ii) the evidence of two origins for the observed amorphous carbon layers: either the ion bombardment amorphisation of the CFC material (erosion zone) or the formation of carbon deposit (thick and thin deposit zones).

Further studies are planned to better elucidate the role of the surface temperature and the plasma processes in the heterogeneity of the deposits.

**Acknowledgment.** We acknowledge the Euratom-CEA association, the Fédération de Recherche FR-FCM, the EFDA European Task Force on Plasma Wall Interactions, the French agency ANR (ANR-06-BLAN-0008 contract) for financial support.

**Table 1**. D content deduced from micro-Raman spectroscopy analysis. (The tile numbering convention is FxTy for tile y on finger x, more details on the position can be found in [5, 13]).

| Tile | Zone | % D (Top) | % D (Gaps) |
|---|---|---|---|
| *F27T5* | *Thick deposit* | 0-23 | 17-27 |
| *F28T5* | *Thick deposit* | 0-22 | - |
| *F17T7* | *Thin deposit* | 17-24 | 17-28 |
| *F10T10* | *Thin deposit* | 14-18 | 21-26 |
| *F27T10* | *Eroded* | 13-20 | 16-20 |
| *F5T3* | *Eroded* | 17-22 | - |

**Figure captions**

**Figure 1**. Raman spectrum obtained for a plasma-deposited hard amorphous carbon layer compared with a spectrum recorded on the top surface of a tile from the thick deposit region. $\sigma_G$, $\Gamma_G$, $I_G$ are the G band wavenumber, full width at half maximum and intensity, respectively, and $\eta$ is the slope of the photoluminescence background.

**Figure 2.** $\sigma_G$ against $\Gamma_G$ plot for tiles coming from the three zones (left side: top surfaces, right side: gap surfaces), compared to a heat treated a-C:H layer. Empty symbols correspond to spectra recorded on the gap surface close to the top surface (for details see text).

**Figure 1**

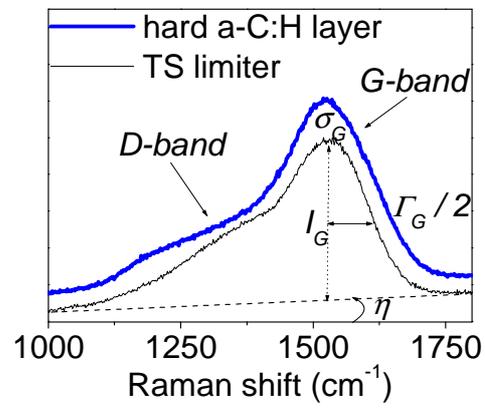

**Figure 2.**

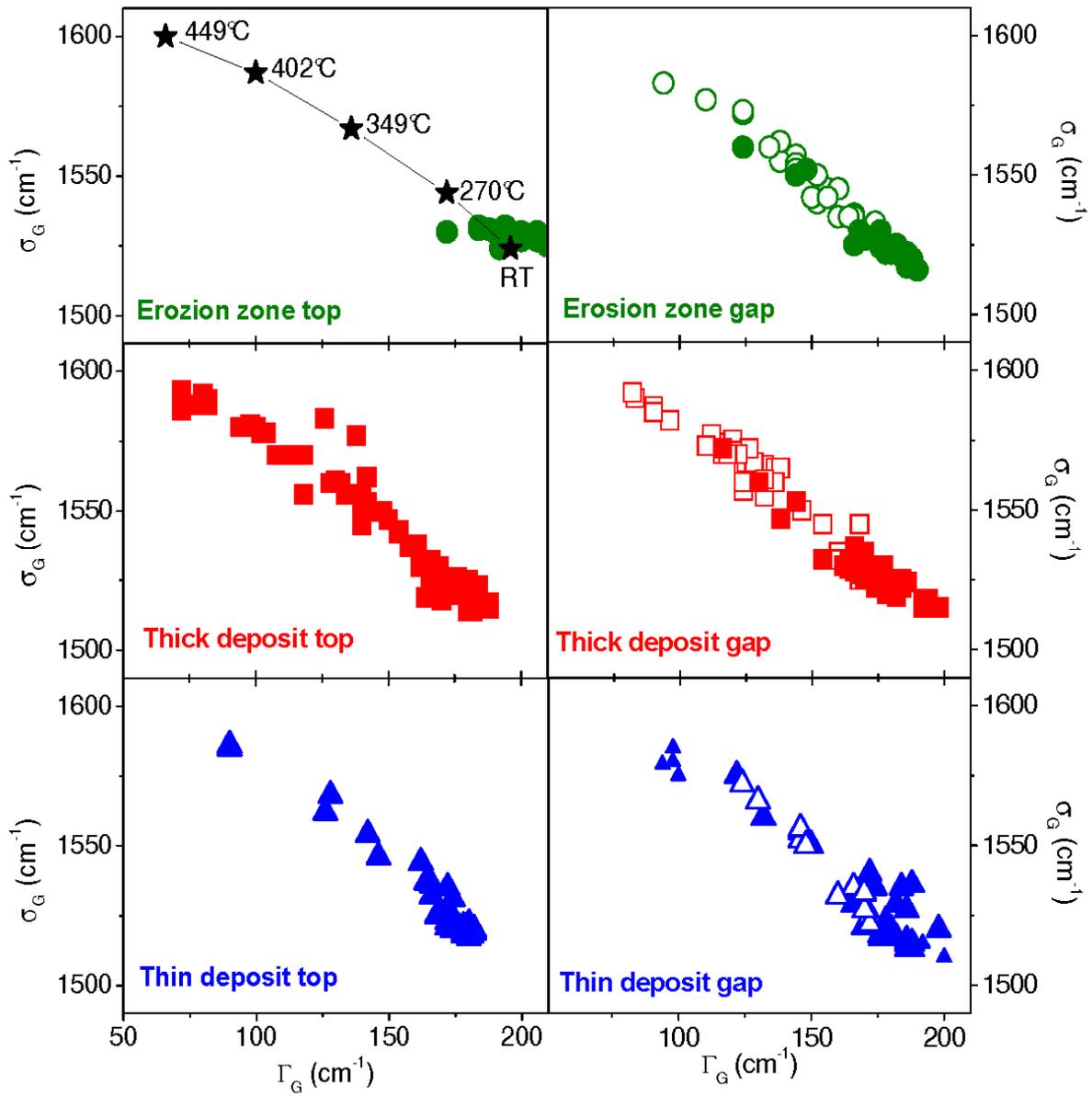